# Deep Learning-Accelerated Dynamic Kinetic Monte Carlo Simulation for Hydrogen Transport in Tungsten

**Seiki Saito [1], Keisuke Takeuchi [1], Hiroaki Nakamura [2,3], Yasuhiro Oda [4], Kazuo Hoshino [5], Yuki Homma [6], Shohei Yamoto [7], Yuki Uchida [8]**

[1] Graduate School of Science and Engineering, Yamagata University, Yonezawa, Japan

[2] Department Research, National Institute for Fusion Science, Toki, Japan

[3] Graduate School of Engineering, Nagoya University, Nagoya, Japan

[4] Simulation Engineering Division, Toyota Technical Development Corporation, Japan

[5] Department of Applied Physics and Physico-Informatics, Keio University, Yokohama, Japan

[6] National Institutes for Quantum Science and Technology, Rokkasho, Japan

[7] National Institutes for Quantum Science and Technology, Naka, Japan

[8] National Institutes of Technology, Nagaoka College, Nagaoka, Japan

* Correspondence: saitos@yz.yamagata-u.ac.jp

**Abstract:** In magnetic confinement fusion reactors, hydrogen plasma irradiation leads to material saturation and recycling, where hydrogen released from the tungsten wall significantly impacts the peripheral plasma. Kinetic Monte Carlo (kMC) simulations are indispensable for investigating the processes leading to a dynamic balance between incident and emitted fluxes at the atomic scale. However, standard kMC frameworks are inadequate for handling realistic material complexities, such as polycrystalline structures and the dynamic evolution of the material under irradiation, as they remain computationally bottlenecked by the need to continuously update transition parameters. To evaluate the migration barriers required for kMC in systems with crystal lattice disorder at grain boundaries (e.g., in polycrystalline materials) or in dynamically evolving structures, one conventional approach is to rely on on-the-fly atomistic calculations using techniques like the Nudged Elastic Band (NEB) method. However, this paper presents a deep learning-accelerated Dynamic kMC framework that eliminates the reliance on such computationally prohibitive calculations. Our approach seamlessly integrates a three-stage deep learning pipeline: a pix2pix model for predicting local 3D potential energy distributions, a U-Net model for extracting hydrogen trapping sites, and a 3D Convolutional Neural Network (3D-CNN) for directly evaluating migration barriers. To achieve macroscopic simulation timescales, we implemented a hierarchical spatial index for event selection combined with a differential local-update algorithm operating in $O(1)$ complexity. This highly optimized architecture restricts parameter recalculation exclusively to the immediate vicinity of moving atoms, accelerating update times. The framework's capability was demonstrated through a large-scale simulation of a realistic polycrystalline tungsten model, successfully reproducing the preferential trapping and accumulation of hydrogen along grain boundaries. Ultimately, this methodology bridges the gap between atomic-scale accuracy and macroscopic timescales for comprehensive full-scale simulations of plasma-wall interactions.

---

## 1. Introduction

Realizing steady-state operation in magnetic confinement fusion reactors relies heavily on managing the severe environment at the plasma-facing materials (PFMs). Tungsten is currently considered the most promising candidate for PFMs, such as the divertor, due to its high melting point, high thermal conductivity, and low sputtering yield. However, during reactor operation, the tungsten wall is continuously bombarded by high fluxes of energetic hydrogen isotopes and helium. This intensive plasma-wall interaction (PWI) [1, 2] induces significant microstructural evolution within the material, including the generation of vacancies, dislocations, and voids. Furthermore, how hydrogen diffuses within a dynamically evolving tungsten matrix containing irradiation-induced defects or growing helium bubbles remains poorly understood. These irradiation-induced defects act as strong trapping sites for hydrogen isotopes. Consequently, the material eventually reaches a state of saturation, driving a fuel recycling process where hydrogen released from the wall significantly impacts the peripheral plasma. Investigating this dynamic evolution, specifically the process until the incident and emitted fluxes reach a dynamic balance at the atomic scale, is essential for evaluating macroscopic tritium retention, ensuring reactor safety, and understanding how recycled hydrogen, including the effects of atoms, molecules, and their vibration-rotation states, impacts the peripheral plasma. Because the transport behavior of hydrogen is intimately coupled with this continuous structural evolution, elucidating this complex spatiotemporal phenomenon remains one of the most formidable challenges in fusion materials science.

To computationally investigate such multiscale phenomena, a hybrid approach combining Molecular Dynamics (MD) and kinetic Monte Carlo (kMC) simulations [3, 4] has emerged as a highly logical framework. In this scheme, MD is utilized to resolve short-timescale, non-equilibrium collisional processes, such as the injection of plasma particles and the formation of collision cascades. Subsequently, the system is handed over to the kMC method, which efficiently simulates the long-timescale, thermally activated diffusion of hydrogen within the bulk material. Despite its theoretical elegance, the practical execution of a true "dynamic"

MD-kMC hybrid simulation has been severely hindered by a critical computational bottleneck. Standard kMC frameworks are inadequate for handling realistic material complexities, such as polycrystalline structures featuring crystal lattice disorder at grain boundaries, or the dynamic evolution of the material under continuous irradiation. In these realistic scenarios, the energy landscape dynamically shifts, requiring the kMC transition parameters to be updated continuously. Conventionally, evaluating the migration barriers required for kMC in such disordered or dynamically evolving structures requires computationally intensive on-the-fly atomistic calculations, such as the Nudged Elastic Band (NEB) method [5]. Performing NEB calculations repetitively for thousands of evolving migration paths during a macroscopic simulation is computationally prohibitive, making real-time dynamic tracking effectively impossible.

To overcome this fundamental limitation, this paper presents a deep learning-accelerated Dynamic kMC framework that eliminates the reliance on such computationally prohibitive on-the-fly atomistic calculations. Our approach seamlessly integrates a three-stage deep learning pipeline. Building upon our previous works, we utilize a pix2pix-based architecture for the rapid prediction of local 3D potential energy distributions (Model-A) [6] , and a U-Net-based model for extracting the precise spatial positions of hydrogen trapping sites (Model-B) [7]. To complete this pipeline, we introduce the final essential component: a 3D Convolutional Neural Network (3D-CNN) that directly predicts migration barriers from local atomic configurations (Model-C) [8].

More importantly, this paper details the architectural integration of these three machine learning models into a unified kMC solver. To achieve macroscopic simulation timescales, we implemented a hierarchical spatial index for event selection combined with a differential local-update algorithm operating in $\boldsymbol{O(1)}$ complexity. This highly optimized architecture restricts parameter recalculation exclusively to the immediate vicinity of moving atoms, accelerating update times from minutes down to the range of hundreds of milliseconds to a few seconds. The framework's capability is demonstrated through a large-scale simulation of a realistic polycrystalline tungsten model, successfully reproducing the preferential trapping and accumulation of hydrogen along grain boundaries. By bridging the gap between atomistic accuracy and macroscopic timescales, this methodology provides a practical pathway for simulating the long-term, dynamic transport of hydrogen isotopes in plasma-facing materials under realistic fusion reactor conditions.

## 2. Methodology

### *2.1. Overall Framework of Deep Learning-Accelerated kMC*

In a standard kinetic Monte Carlo (kMC) simulation, the system evolves from state to state based on pre-calculated transition rates. However, under plasma irradiation, the energy landscape is continuously modified due to the dynamic evolution of the atomic structure. To track this dynamic evolution without relying on computationally prohibitive on-the-fly atomistic calculations, we developed a highly localized, deep learning-based dynamic prediction framework.

The core philosophy of our approach relies on the principle of locality: the transition behavior of a hydrogen (H) atom is predominantly governed by its immediate local atomic environment rather than the macroscopic state of the entire bulk. First, the simulation domain is discretized into a 3D grid with a uniform voxel size of 0.1 Å. Instead of evaluating the entire system simultaneously, the algorithm extracts a localized 3D voxel space centered on each individual H atom. Before entering the kMC time-evolution loop, a three-stage deep learning pipeline is executed for all initially distributed H atoms to evaluate their available trapping sites and corresponding migration barriers. Crucially, as the simulation progresses, this same localized pipeline is dynamically triggered to update these parameters strictly where atomic transitions occur, enabling highly efficient on-the-fly tracking.

### *2.2. Deep Learning Pipeline for Local Transition Parameters*

For each H atom in the system, the transition parameters are evaluated through the following sequential pipeline, integrating our previously developed surrogate models, as illustrated in Figure 1.

#### *2.2.1 Prediction of Local 3D Potential Distributions (Model-A)*

A local volumetric region of 12.8×12.8×12.8 Å (128×128×128 voxels) centered on the target H atom is extracted. The atomic coordinates of tungsten (W) and H within this region are converted into a two-channel tensor. This tensor is fed into a pix2pix-based neural network (Model-A) to rapidly predict the local 3D binding energy distribution, $U_{bind}(\boldsymbol{r})$. The ground truth for this prediction is based on the Embedded Atom Method (EAM) potential [9] developed for tungsten-hydrogen interactions. The total potential energy $U$ of the system, which forms the basis for this calculation, is defined as:

$$U = \sum_{i} E_i^{\text{emb}} \left[ \sum_{j \neq i} \rho_j(r_{ij}) \right] + \sum_{i,j>i} \phi_{i,j}(r_{ij})$$

where $E_i^{\text{emb}}$ is the embedding energy of atom $i$, $\rho_j$ is the effective electron density contributed by atom $j$, $\phi_{i,j}$ is the pairwise interatomic potential, and $r_{ij}$ is the distance between atoms $i$ and $j$. By utilizing this potential function, the many-body effects of metallic bonding are effectively incorporated, allowing for an accurate evaluation of the energy landscape associated with particle migration.

#### *2.2.2 Extraction of Hydrogen Trapping Sites (Model-B)*

The predicted potential distribution is subsequently processed by a U-Net-based convolutional neural network (Model-B), which outputs a continuous probability map indicating stable trapping sites for the central H atom. To deterministically extract physically valid, discrete trapping sites from this continuous map, a Non-Maximum Suppression (NMS) algorithm is applied. By setting a probability threshold of 0.5 and scanning the 26-neighboring voxels within the physical jump distance, local maxima are isolated. This procedure comprehensively identifies the set of all possible target sites to which the central H atom can migrate.

*2.2.3 Direct Prediction of Migration Barriers (Model-C)*

Once the initial position $\boldsymbol{r}_i$ and the candidate target sites $\boldsymbol{r}_j$ are identified, the migration barrier $\Delta U_{i\to j}$ for each path is evaluated using a 3D-CNN (Model-C). The input to this model is a localized 63×63×63 voxel space containing two channels: (1) the cropped 3D potential energy distribution, and (2) a binary map flagging only the spatial coordinates of the initial and final states. The model extracts the topological features of the energy landscape and directly outputs the scalar migration barrier $\Delta U_{i\to j}$, entirely bypassing iterative Nudged Elastic Band (NEB) calculations.

Based on the predicted barrier, the transition rate $R_{i\to j}$ is calculated using the Arrhenius equation:

$$R_{i\to j} = R_0 \exp\left(-\frac{\Delta U_{i\to j}}{k_\mathrm{B}T}\right)$$

where $k_\mathrm{B}$ is the Boltzmann constant and $T$ is the absolute temperature. The attempt frequency $R_0$ is defined as $R_0 = 6D_0/(\Delta x^2)$, assuming the average jump distance $\Delta x = a/2$ for the bcc tungsten lattice and $D_0$ as the pre-exponential factor of the diffusion coefficient. In this study, the simulations were performed at a temperature of $T = 300$ K. $D_0$ was set to $4 \times 10^{13}$ Å$^2/s$. With a tungsten lattice constant of $a = 3.16$ Å (yielding an average jump distance of $\Delta x = 1.58$ Å), the attempt frequency $R_0$ was evaluated to be approximately $9.6 \times 10^{13}\ s^{-1}$.

*2.3 Hierarchical Spatial Grid for Fast Event Selection*

Executing the pipeline described above for all H atoms yields a comprehensive list of transition rates. In large-scale simulations, storing these rates in a simple 1D array and performing a linear search ($O(N)$ complexity) to select the next event becomes a severe computational bottleneck. To accelerate the event selection process, we implemented an optimized algorithm based on a spatial extension of the Bortz-Kalos-Lebowitz (BKL) method [10], which is widely known as the n-fold way.

The simulation box is divided into microscopic 3D cells. The transition rates are aggregated and stored in a three-tier hierarchical index array representing the sum of rates in cells (points), lines, and surfaces:

1. Point (Cell) Rate: $R_\mathrm{point}(x, y, z)$
2. Line Rate: $R_\mathrm{line}(y, z) = \sum_x R_\mathrm{point}(x, y, z)$

3. Surface Rate: $R_{\text{surface}}(z) = \sum_y R_{\text{line}}(y, z)$

The total transition rate of the system is $\Gamma = \sum_z R_{\text{surface}}(z)$. The event selection process utilizes a two-step random number scheme. First, a random number $u_1 \in (0,1]$ is generated to identify the moving atom. The target cumulative rate $R_{\text{target}} = u_1 \cdot \Gamma$ is used to systematically drill down the hierarchy: identifying the active $z$-surface, then the $y$-line, the specific $(x, y, z)$-cell, and finally the exact H atom within that cell. Once the moving atom is uniquely identified, a second, independent random number $u_2 \in (0,1]$ is generated to select the specific target trapping site among the available transition paths for that atom.

*2.4 Local Update Algorithm for Dynamic Tracking*

When an H atom executes a move, the local atomic configuration changes, altering the potential energy landscape for the moved atom and its neighbors. Recalculating the entire system's parameters at every kMC step is computationally inefficient. Instead, we introduce a highly efficient local update scheme, as schematically represented in Figure 2.

Because the deep learning models only require a localized spatial input (up to 12.8 Å), the parameter updates can be strictly confined to the immediate vicinity of the transition event. Upon an atom's movement, the algorithm flags only the "moved H atom" and any other H atoms residing within a 3×3×3 cell neighborhood surrounding both the initial and final positions. The deep learning pipeline (Model-A to C) is re-executed exclusively for these flagged atoms, with their individual inference tasks processed in parallel across multiple GPUs.

The hierarchical index arrays are then updated using a delta-update logic. For an updated atom $i$ located in cell $(x, y, z)$, the old sum of its transition rates $R_{\text{old}}^{(i)}$ is subtracted, and the newly evaluated sum $R_{\text{new}}^{(i)}$ is added:

$$R_{\text{point}}(x, y, z) \leftarrow R_{\text{point}}(x, y, z) - R_{\text{old}}^{(i)} + R_{\text{new}}^{(i)}$$

$$R_{\text{line}}(y, z) \leftarrow R_{\text{line}}(y, z) - R_{\text{old}}^{(i)} + R_{\text{new}}^{(i)}$$

$$R_{\text{surface}}(z) \leftarrow R_{\text{surface}}(x, y, z) - R_{\text{old}}^{(i)} + R_{\text{new}}^{(i)}$$

$$\Gamma \leftarrow \Gamma - R_{\text{old}}^{(i)} + R_{\text{new}}^{(i)}$$

This differential update mechanism reduces the computational complexity of the array maintenance from $O(N)$ to a constant time $O(1)$. This architectural design is what ultimately enables the deep learning models to be seamlessly integrated into the kMC solver, realizing a true Dynamic kMC simulation at practical computational speeds.

**3. Results and Discussion**

*3.1 Setup of the Polycrystalline Tungsten Model*

To demonstrate the capability of the developed deep learning-accelerated Dynamic kMC framework, we performed a simulation of hydrogen atom transport in a realistic tungsten microstructure. Instead of a simple perfect crystal, a polycrystalline tungsten structure was generated using the polypal code [11]. Based on the simulation configuration, the computational domain was set to a $63.2 \times 63.2 \times 63.2$ Å$^3$ box with periodic boundary conditions applied in all three spatial directions ($x$, $y$, and $z$). The structure consists of uniform grains with an average grain diameter of 50 Å, introducing complex grain boundaries (GBs) with a clearance of 3.0 Å. The tungsten lattice parameter was set to $a = 3.16$ Å.

Following the construction of the polycrystalline tungsten lattice, hydrogen atoms were initially introduced at random positions throughout the simulation box. Considering the high hydrogen inventory typically observed in tungsten under severe fusion-relevant plasma irradiation, the hydrogen-to-tungsten (H/W) ratio was set to 0.2. The H/W ratio of 0.2 was selected based on the concept of dynamic retention during plasma irradiation. While post-irradiation measurements of tungsten typically report lower H/W values, the in-situ concentration [12, 13] during active plasma bombardment can be significantly higher due to the dynamic balance between incident and recycled hydrogen fluxes, justifying the use of H/W=0.2 as a representative bulk concentration in this kMC framework. This relatively high-concentration setting allows us to capture complex diffusion behaviors, such as trap saturation and the preferential accumulation of hydrogen at grain boundaries. To eliminate unphysical overlapping of atoms and to settle the hydrogen atoms into their initial local minimum energy states (trapping sites), an initial energy minimization (structural relaxation) was performed. This relaxed state served as the starting configuration (time t=0) for the subsequent Dynamic kMC simulation.

*3.2. Hydrogen Transport Trajectories and Trapping at Grain Boundaries*

Starting from the energy-minimized configuration, the long-term transport of hydrogen (H) atoms was simulated using the proposed Dynamic kMC algorithm. Figure 3 illustrates the sequential stages of the simulation setup and the resulting transport dynamics. Figure 3(a) displays the complex polycrystalline tungsten (W) structure generated by the polypal code, featuring well-defined grain boundaries surrounding uniform grains with an average diameter of 50 Å. The initial spatial distribution of H atoms immediately after the relaxation process is presented in Figure 3(b), where H atoms (red spheres) have settled into the local potential minima of the W lattice. Figure 3(c) illustrates the final spatial positions of the H atoms at the end of the simulation, vividly capturing their significant segregation and accumulation along the grain boundaries.

Figure 3(d) shows the accumulated trajectories of the H atoms throughout the simulation. A clear distinction in diffusion behavior is evident between the bulk crystal and the grain boundary regions. In the crystalline bulk regions, H atoms exhibit typical interstitial random-walk behavior characterized by linear and rapid jump sequences. However, upon encountering a grain boundary, the trajectories change drastically, becoming densely clustered along the boundary planes. This evidence directly demonstrates that H atoms are effectively captured by the deep potential wells formed at the GBs. Moreover, the trajectories indicate that the GBs not only act as strong trapping sites but also serve as low-dimensional pathways for preferential migration (grain boundary diffusion) under realistic multi-grained environments.

*3.3. Quantitative Correlation between Grain Boundary Proximity and Residence Time*

To rigorously validate the physical consistency of our deep learning-accelerated framework, we performed a statistical analysis correlating the spatial proximity of H atoms to the nearest grain boundary with their respective residence times at each trapping site. The grain boundaries were mathematically defined as Voronoi interfaces derived from the nearest and second-nearest grain seeds. Figure 4 highlights the quantitative results obtained from this correlation analysis. Figure 4(a) displays a violin plot of the log-scaled residence time as a function of the spatial distance from the nearest grain boundary, with the red markers indicating the median values. A sharp, continuous increase in the median residence time is observed as the distance approaches zero. In the bulk regions (distance > 5 Å), the residence time remains short and tightly distributed, reflecting the low and uniform migration barriers characteristic of the perfect bcc lattice. Conversely, within the immediate vicinity of the GBs (distance < 2 Å), the residence time distribution extends upward by several orders of magnitude, and the median value shifts significantly higher.

This statistical trend is further substantiated by the conditional probability analysis shown in Figure 4(b). The plot evaluates the probability $P(t > 90\text{th percentile})$, representing the likelihood of an H atom undergoing an exceptionally long trapping event (top 10% of all residence times across the simulation), plotted against the GB distance. While the baseline probability is expected to be 10% in the random bulk environment, it monotonically scales up to a remarkably higher percentage as the distance to the GB decreases. This highly localized trapping effect is statistically significant, as confirmed by a Spearman rank correlation analysis ($r = -0.046,\ p = 4.16 \times 10^{-20}$). The relatively small absolute value of the correlation coefficient naturally arises from the physical nature of the system: the prolonged residence time is strictly confined to the immediate vicinity of the grain boundaries, whereas hydrogen mobility in the vast bulk region remains completely unconstrained and independent of the boundary distance.

From a physical standpoint, these statistics provide definitive proof that our integrated three-stage deep learning pipeline operates with high physical fidelity. Because the residence time $\Delta t$ in the BKL algorithm is inversely proportional to the transition rate ($\Delta t \propto 1/R$), these prolonged residence times mean that the 3D-CNN (Model-C) successfully outputs significantly higher migration barriers $\Delta U$ when an H atom attempts to escape from a GB site. This high-barrier prediction is inherently driven by the deep 3D binding energy distributions predicted by Model-A and the high-density trapping site networks mapped by Model-B at the defective GB interfaces. The ability to accurately reproduce such macro-statistical physical phenomena directly from a chain of neural network inferences underscores the robustness of this surrogate-based approach for multiscale plasma-wall interaction modeling.

*3.4 Evaluation of the Dynamic kMC Framework*

To quantitatively evaluate the computational efficiency and practical feasibility of the proposed deep learning-accelerated Dynamic kMC framework, we monitored the execution time for the transition parameter updates and compared it against the conventional approach. Figure 5 highlights the execution time benchmark results under varying system scales (i.e., the total number of hydrogen atoms in the system). In a conventional on-the-fly kMC simulation, whenever an atomic displacement occurs, the minimum energy paths and migration barriers for all possible transition events must be recalculated using atomistic methods such as the Nudged Elastic Band (NEB) method. As illustrated by the dashed baseline in Figure 5, the computational cost of the conventional NEB-based approach scales heavily with the size of the system and the number of active diffusion paths, requiring several minutes to hours per single kMC step. Such a prohibitive increase in computational time severely restricts the total simulation timescale, making it practically impossible to observe macroscopic hydrogen transport phenomena in larger, complex microstructures like polycrystals.

In stark contrast, our framework achieves an unprecedented speedup, reducing the average update time per kMC step to the millisecond order, as demonstrated in Figure 5. Crucially, the distribution of execution times recorded over 291,194 steps confirms that the per-step cost remains stable throughout the entire simulation run, with the majority of steps completing within sub-second timescales. This temporal stability provides empirical evidence of the $O(1)$ complexity introduced by the differential local-update algorithm.

By isolating the computational domain to an immediate 3×3×3 cell neighborhood surrounding the transition event, the deep learning pipeline (Model-A to Model-C) only re-evaluates a small number of localized atoms. This remarkable efficiency and robustness enabled us to successfully execute a long-term simulation spanning a total of 291,194 kMC steps, which was completed within a total wall-clock time of approximately 24 hours. This massive calculation was performed by utilizing four AMD Instinct MI300A Accelerated Processing Units (APUs) in parallel on Subsystem B of the Plasma Simulator supercomputer system, where each APU highly integrates 24 "Zen 4" CPU cores and a CDNA 3 GPU architecture with 228 Compute Units. Throughout this extensive run, the simulation proceeded smoothly without encountering any numerical freezing or performance degradation, demonstrating the framework's capability to stably track hydrogen transport dynamics within the complex polycrystalline lattice.

## 4. Conclusion

In this study, we have successfully developed, implemented, and validated a deep learning-accelerated Dynamic kinetic Monte Carlo (kMC) framework designed to simulate long-term hydrogen (H) isotope transport in complex tungsten (W) microstructures. By sequentially coupling three specialized neural network architectures, which predict local 3D binding energy landscapes (Model-A), map discrete trapping site networks via a U-Net and Non-Maximum Suppression (NMS) algorithm (Model-B), and directly evaluate saddle-point migration barriers via a 3D-CNN (Model-C), we have entirely bypassed the prohibitive computational cost of conventional on-the-fly atomistic methods such as the Nudged Elastic Band (NEB) technique. The structural and computational integrity of this framework relies on an optimized spatial tracking architecture. By executing a hierarchical spatial grid for event selection and a differential local-update scheme, the parameter recalculation is strictly isolated within a 3×3×3 cell neighborhood surrounding each migration event. Benchmark performance data empirically validated that this algorithmic design achieves a constant-time computational complexity of $O(1)$, stabilizing the transition parameter update time to the millisecond order regardless of the overall system scale or the total number of active H solutes.

The practical capability and physical fidelity of this framework were demonstrated through a large-scale simulation within a realistic polycrystalline tungsten microstructure generated via the polypal code. The integrated solver smoothly executed a long-term diffusion run spanning a total of 291,194 steps (requiring approximately 24 hours of computational time utilizing four AMD Instinct MI300A APUs on Subsystem B of the Plasma Simulator), successfully tracking the dynamic transport behavior without suffering from numerical freezing or performance degradation. Furthermore, a rigorous statistical analysis of the simulation trajectory successfully captured a highly localized trapping effect at the GBs. While the global Spearman rank correlation coefficient between grain boundary proximity and log-scaled residence time was small ($r = -0.046$) due to the unconstrained, independent random-walk behavior dominant in the vast crystalline bulk, the analysis demonstrated a mathematically robust and statistically significant correlation ($p = 4.16 \times 10^{-20}$). This statistical trend quantitatively proves that the multi-stage neural network inferences consistently dictate physically reasonable, prolonged residence times specifically confined to the immediate vicinity (distance < 2 Å) of defective GB interfaces.

In conclusion, this methodology successfully bridges the multiscale gap in plasma-wall interaction (PWI) modeling, delivering atomistic-level energy landscape precision at long kMC timescales. Future enhancements of this framework will expand the surrogate training datasets to incorporate more complex irradiation-induced microstructural defects, such as self-interstitial clusters, multi-vacancy voids, dislocations, and helium bubbles. By linking this advanced Dynamic kMC solver with direct Molecular Dynamics (MD) simulations for the initial plasma implantation phases, we aim to establish a comprehensive, full-scale multiscale predictive modeling paradigm for fuel recycling and tritium retention in future magnetic confinement fusion reactors.

**Acknowledgement**

The Dynamic kMC framework presented in this study is based on the methodology developed under Grant-in-Aid for Scientific Research No. 22K03572 from the Japan Society for the Promotion of Science, Japan, and the NIFS Collaborative Research Program (NIFS25KSPT009, NIFS24KIPT013, NIFS25KIST066, and NIFS22KIGS002). S. Saito is supported by JST (Moonshot R&D Program) Japan Grant Number JPMJMS24A3 for the simulation parameter settings and validation conducted in accordance with the wall model requirements.

The computations were performed using the Plasma Simulator of NIFS (Toki, Gifu, Japan).

The authors would like to acknowledge the use of Gemini (Gemini 3 Flash, Google) for assistance in English language editing and the structural organization of this manuscript.

## References

[1] G. Janeschitz, K. Borrass, G. Federici, Y. Igitkhanov, A. Kukushkin, H. D. Pacher, G. W. Pacher, and M. Sugihara, "The ITER divertor concept," J. Nucl. Mater., vol. 220-222, pp. 73-87, 1995.

[2] G. F. Matthews, "Plasma detachment from divertors and limiters," J. Nucl. Mater., vol. 220-222, pp. 104-116, 1995.

[3] K. O. E. Henriksson, K. Nordlund, A. Krasheninnikov, and J. Keinonen, "Differences in the defect structure of ion-irradiated carbon nanotubes and graphene," Appl. Phys. Lett., vol. 87, no. 16, p. 163113, 2005.

[4] A. M. Ito et al., "Kinetic Monte Carlo simulation of dynamics of hydrogen-vacancy clusters in tungsten," Nucl. Fusion, vol. 55, no. 7, p. 073013, 2015.

[5] G. Henkelman, B. P. Uberuaga, and H. Jónsson, "A climbing image nudged elastic band method for finding saddle points and minimum energy paths," J. Chem. Phys., vol. 113, no. 22, pp. 9901-9904, 2000.

[6] S. Sato, H. Nakamura, C. Takahashi, K. Sawada, K. Hoshino, M. Kobayashi, and M. Hasuo, "Deep learning model for predicting the spatial distribution of binding energy from atomic configurations," Jpn. J. Appl. Phys., vol. 63, no. 9, p. 09SP03, 2024.

[7] S. Saito, K. Takeuchi, H. Nakamura, Y. Oda, K. Sawada, K. Hoshino, Y. Homma, S. Yamoto, and Y. Uchida, "Deep Learning of Hydrogen Trapping Sites in Tungsten for Atomistic Plasma-Wall Simulations," Plasma and Fusion Res., vol. 21, p. 2403017, 2026.

[8] S. Saito, K. Takeuchi, H. Nakamura, Y. Oda, K. Hoshino, Y. Homma, S. Yamoto, and Y. Uchida, "Development of a 3D-CNN-based Prediction Model for Migration Barriers in Plasma-Wall Interactions," Submitted to IEEE Transactions on Plasma Science.

[9] L. F. Wang, X. Shu, G. H. Lu, and F. Gao, "A new analytical interatomic potential for tungsten-hydrogen system," J. Phys. Condens. Matter, vol. 29, no. 43, p. 435401, 2017.

[10] A. B. Bortz, M. H. Kalos, and J. L. Lebowitz, "A new algorithm for Monte Carlo simulation of Ising spin systems", J. Comp. Phys., 17(1), 10-18, 1975.

[11] Y. Shin, V. Moul, K. Kang, and B. Lee, "PolyPal: A parallel microscale virtual specimen generator," Comput. Phys. Commun., 308, 109458, 2025.

[12] M. Yamagiwa, Y. Nakamura, N. Matsunami, N. Ohno, S. Kajita, M. Takagi, M. Tokitani, S. Masuzaki, A. Sagara, and K. Nishimura, "In situ measurement of hydrogen isotope retention using a high heat flux plasma generator with ion beam analysis," Phys. Scr., T145, 014032, 2011.

[13] S. Saito, H. Nakamura, K. Sawada, M. Kobayashi, G. Kawamura, T. Sawada, and H. Masahiro, "Molecular dynamics simulation for hydrogen recycling on tungsten divertor for neutral transport analysis," Jpn. J. Appl. Phys., 60, SAAB08, 2021.

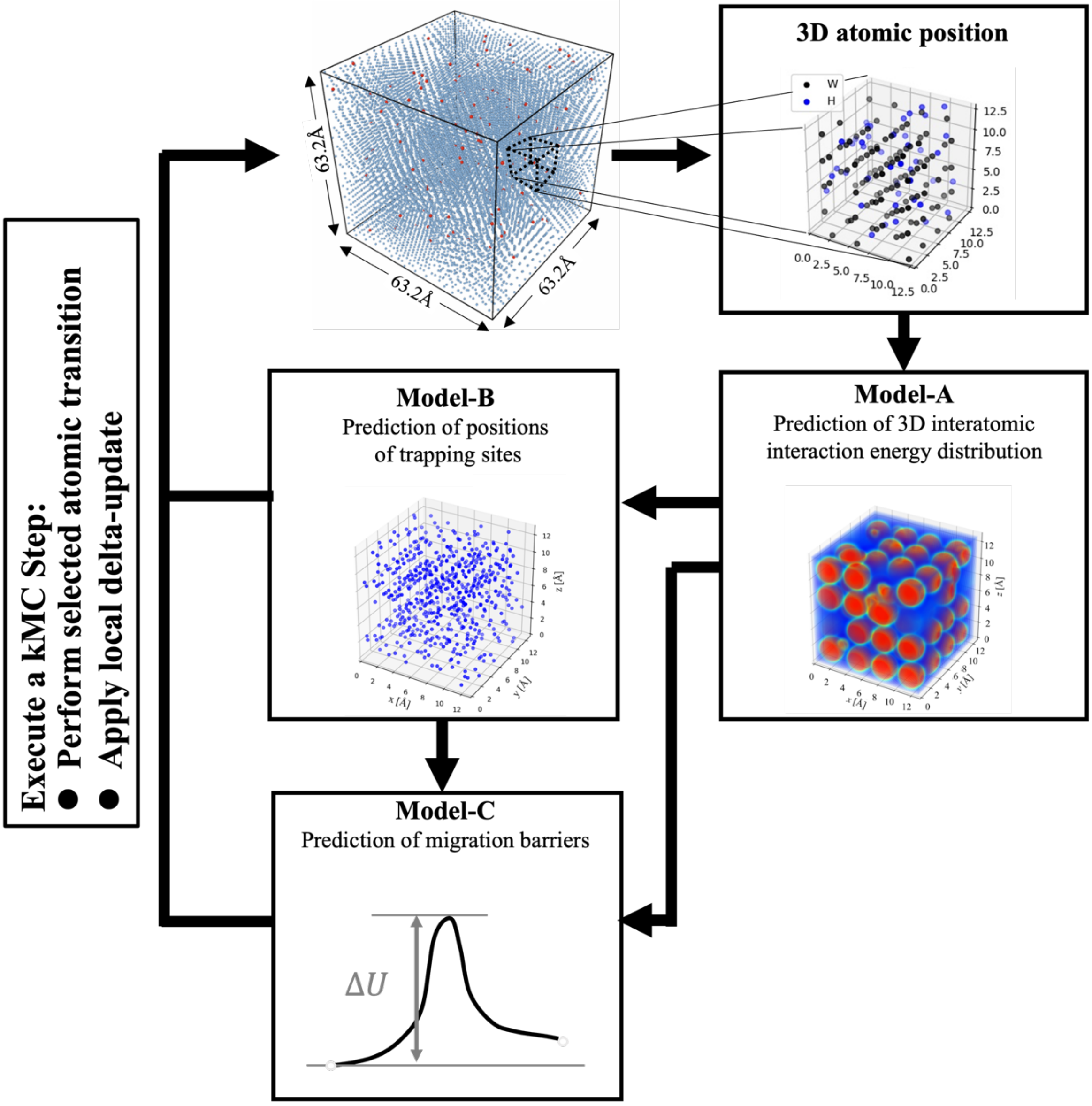

**Figure 1.** Overall architecture of the deep learning pipeline for on-the-fly kMC parameter prediction. A localized spatial region is extracted from the global macroscopic simulation box ($63.2 \times 63.2 \times 63.2$ Å$^3$). Based on this local atomic configuration, Model-A predicts the 3D potential energy distribution, Model-B identifies the spatial positions of accessible trapping sites, and Model-C calculates the migration barrier ($\Delta U$) for the transition events. This localized inference scheme enables rapid and continuous parameter updates during dynamic structural evolution.

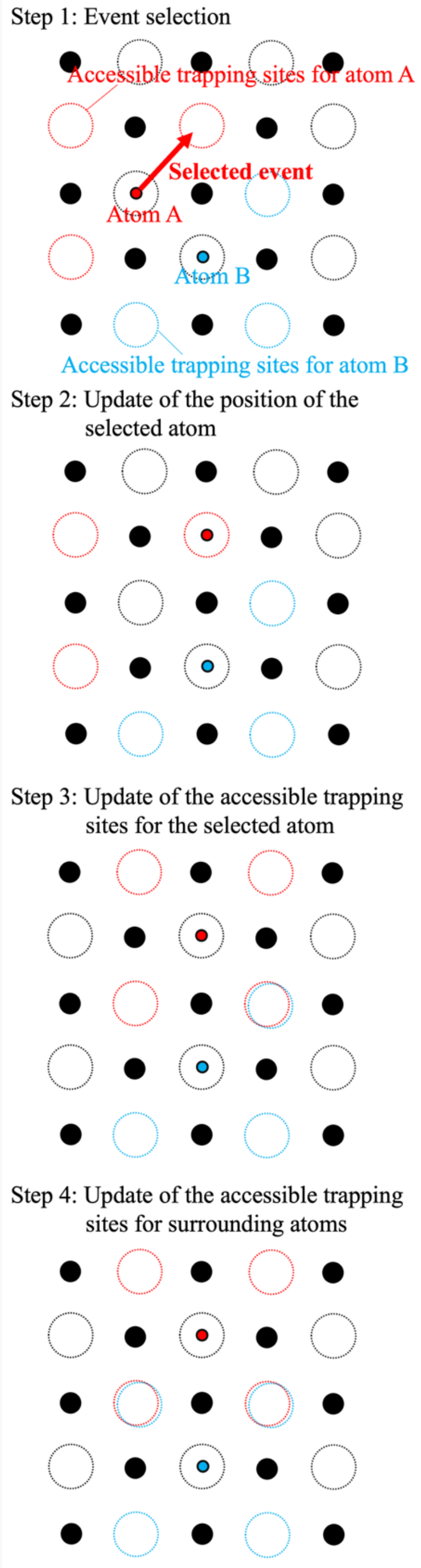

**Figure 2.** Schematic representation of the local update algorithm for dynamic tracking of hydrogen transport. To minimize computational cost, parameter recalculations are strictly confined to the vicinity of a transition event. The sequence proceeds in four steps: (Step 1) selection of a transition event for a specific atom (Atom A); (Step 2) positional update of the selected atom; (Step 3) recalculation of the accessible trapping sites and corresponding migration barriers for the moved atom at its new position; and (Step 4) execution of equivalent updates for the neighboring atoms (e.g., Atom B) whose local potential landscapes were altered by Atom A's movement. This delta-update mechanism ensures optimal computational efficiency.

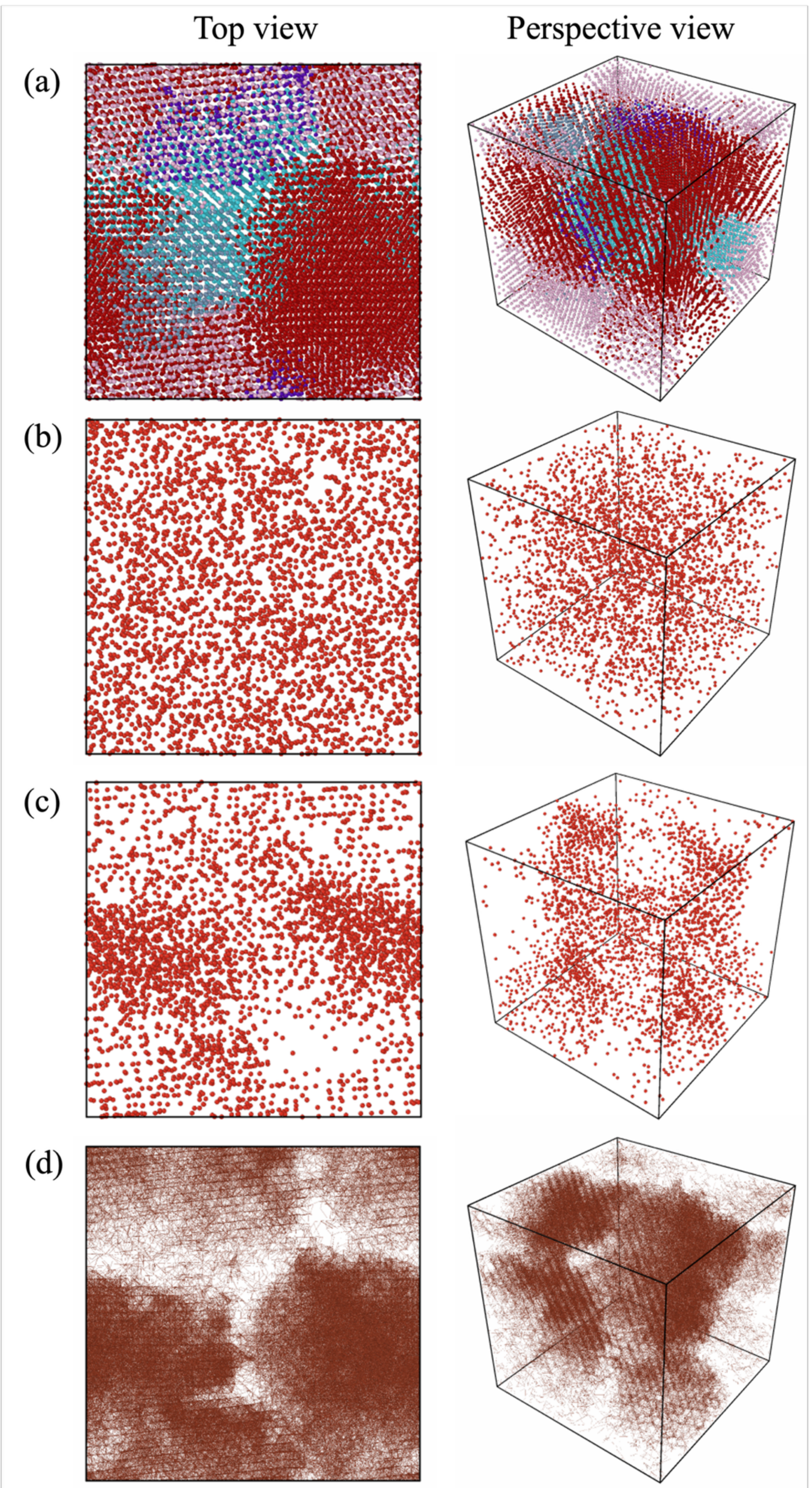

**Figure 3.** Visualization of the Dynamic kMC simulation in a polycrystalline tungsten model. (a) The polycrystalline tungsten lattice generated by the polypal code, featuring multiple grains and complex grain boundaries. (b) Initial spatial distribution of hydrogen atoms (red spheres) after structural relaxation. (c) Final spatial positions of hydrogen atoms at the end of the simulation, showing significant segregation and accumulation along the grain boundaries. (d) Accumulated migration trajectories of hydrogen atoms during the dynamic simulation, which clearly demonstrate random interstitial diffusion within the crystalline bulk followed by preferential trapping and accumulation along the grain boundary interfaces.

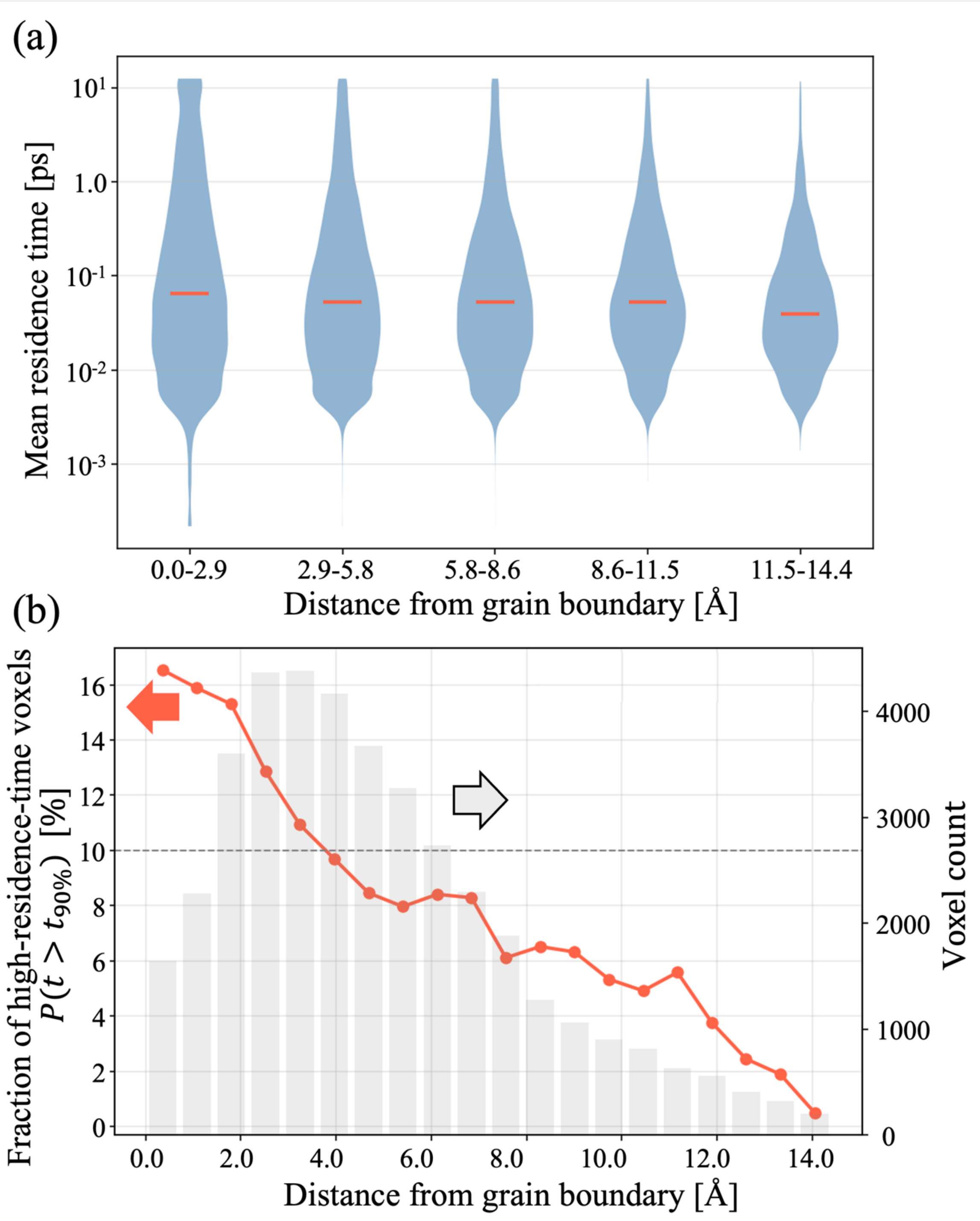

**Figure 4.** Quantitative analysis of the highly localized hydrogen trapping effect at grain boundaries. (a) Violin plot illustrating the distribution of log-scaled residence times as a function of distance from the nearest grain boundary. Red markers denote the median values, demonstrating a sharp increase in trapping duration specifically within the immediate vicinity of the boundaries (distance < 2 Å). (b) The conditional probability of encountering an exceptionally long trapping event (residence time exceeding the 90th percentile) plotted against the distance to the grain boundary. The elevated probability near the interfaces rapidly decays to the expected random baseline (10%) in the bulk region, statistically validating the physical fidelity of the deep learning-predicted migration barriers ($p = 4.16 \times 10^{-20}$).

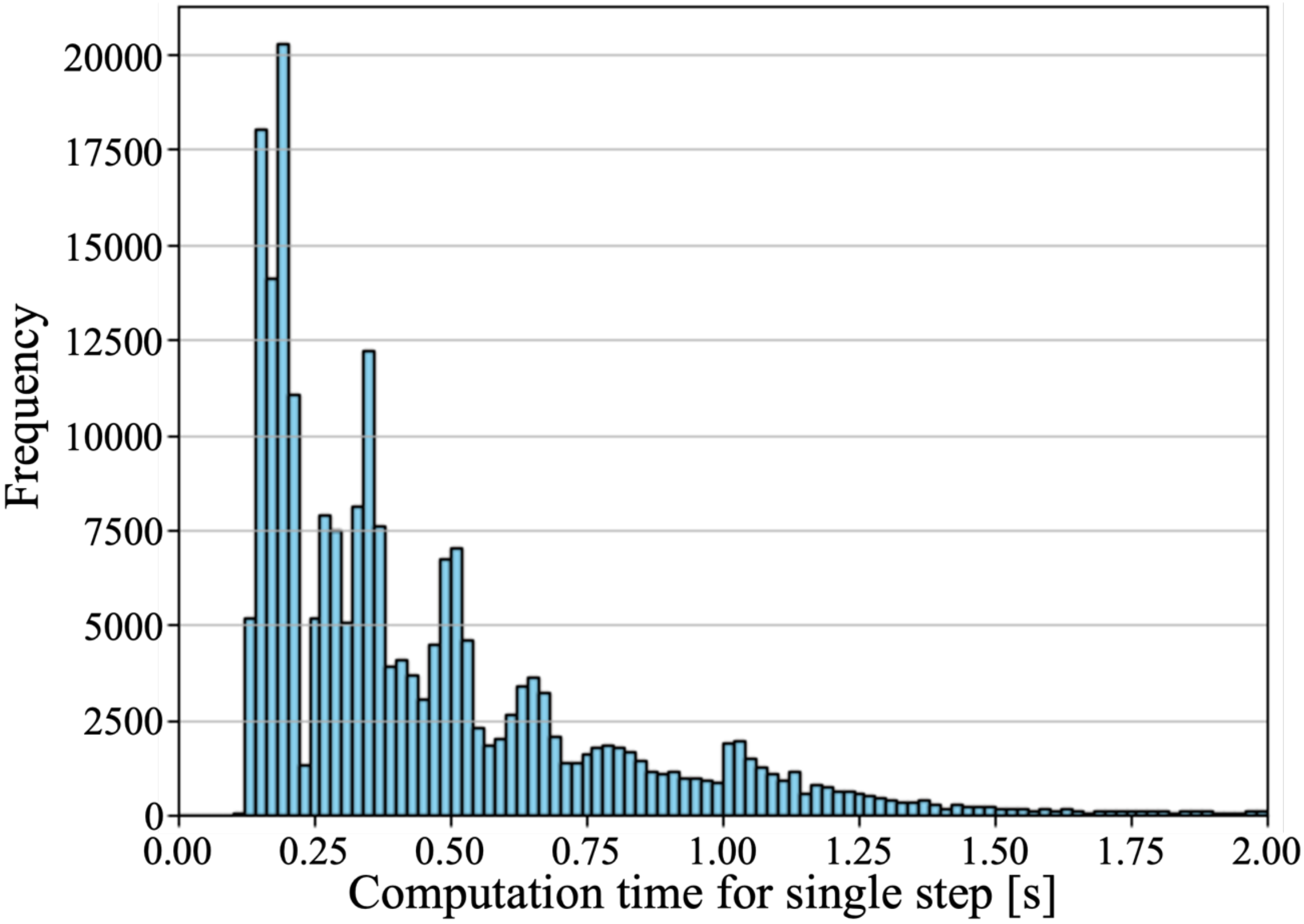

**Figure 5.** Distribution of computational execution times per kMC step recorded over the entire simulation run (291,194 steps). The histogram demonstrates that the vast majority of steps are completed within approximately 0.25–0.50 seconds, with no systematic increase over the course of the simulation.